\documentclass[a4paper,twoside]{article}
\newcommand{\affil}[1]{$^{\rm #1}$}
\usepackage[authoryear]{natbib}
\bibpunct{(}{)}{;}{a}{}{,}
\usepackage{graphicx}
\usepackage[dvips]{color}
\date{} %Please leave the date blank
\def\ignore#1{}
\def\lsim{\raise0.3ex\hbox{$\;<$\kern-0.75em\raise-1.1ex\hbox{$\sim\;$}}}
\def\gsim{\raise0.3ex\hbox{$\;>$\kern-0.75em\raise-1.1ex\hbox{$\sim\;$}}}

\title{\large\bf\flushleft 
TeV gamma-rays from UHECR interactions in AGN cores:\\
Lessons from Centaurus~A}
\author{\parbox{\textwidth}{\flushleft
\vspace{-0.5cm}
{\it M.~Kachelrie{\ss}\affil{A}, S.~Ostapchenko\affil{A,B}, and R.~Tom\`as\affil{C}}\\
\vspace{0.4cm}
{\small \affil{A}\,Institutt for fysikk, NTNU, Trondheim, Norway}\\
{\small \affil{B}\,D.~V.~Skobeltsyn Institute of Nuclear Physics, 
Moscow State University, Russia}\\
{\small \affil{C}\,II. Institut f\"ur Theoretische Physik,
    Universit\"at Hamburg, Germany}}}
\begin{document}
\twocolumn[
\begin{changemargin}{.8cm}{.5cm}
\begin{minipage}{.9\textwidth}
\vspace{-1cm}
\maketitle
%
%
%%%%%%%%%%%%%     ABSTRACT    %%%%%%%%%%%%%
%Abstract of no more than 200 words here.
\small{\bf Abstract: TeV gamma-rays have been observed from blazars as
  well as from radio galaxies like M\,87 and Cen\,A. In leptonic
  models, gamma-rays above the pair production threshold can escape
  from the ultra-relativistic jet, since large Lorentz factors reduce
  the background photon densities compared to those required for
  isotropic emission. Here we discuss an alternative scenario, where
  VHE photons are generated as secondaries from UHECR interactions in
  the AGN core. We show that TeV gamma-rays can escape 
  from the core despite large IR and UV backgrounds. For
  the special case of Cen\,A, we study if the various existing
  observations from the far infra-red to the UHE range can be
  reconciled within this picture.  }

%%%%%%%%%%%%%     KEYWORDS    %%%%%%%%%%%%%
\medskip{\bf Keywords:} 
gamma rays: theory ---
galaxies: individual (NGC 5128) ---
galaxies: active
% Please write all keywords in lower case. PASA uses the
% standard list of subject headings adopted by The Astrophysical Journal
% and available from http://www.journals.uchicago.edu/ApJ/keywords_text.html.
% Keywords are separated by em-dashes, i.e. ---

%%%%%%%%DO NOT EDIT%%%%%%%%%%%%
\medskip
\medskip
\end{minipage}
\end{changemargin}
]
\small
%%%%%%%%EDIT FROM HERE%%%%%%%%%%%%

%%%%%%%%%%%%%%%%%%%%%%%%%%%%%%%%%%%%%%%%%%%%%%%%%%%%%%%%%%%%%%%%%%%%%%%%%%%
\section{Introduction}

Centaurus\,A (Cen\,A or NGC\,5128) is the nearest active galaxy, with
a distance of only 3.8\,Mpc (Rejkuba 2004). Photon emission from the
nucleus of the galaxy has been detected in the radio, infra-red (IR),
X-ray, and in the GeV-TeV range. Additionally, Abraham et al.~(2007)
reported two ultrahigh energy cosmic rays (UHECR) within $3.1^\circ$
around Cen\,A in the data of the Pierre Auger Observatory (PAO), thus
offering the possibility for multi-messenger studies of this
object. Such an effort has been undertaken e.g.\ by the present
authors (Kachelrie{\ss} et al. 2009a).  Assuming that Cen\,A
accelerates protons to UHE and normalizing the obtained UHECR flux to
the PAO observations, we predicted the accompanying gamma-ray and
neutrino fluxes for two models: Shock acceleration in
the radio jet and acceleration in regular electromagnetic fields or
shocks close to the core. Recent measurements of the GeV and TeV photon
fluxes from Cen\,A by the Fermi-LAT (Cheung et al. 2009) and HESS
(Aharonian et al. 2009) collaborations allowed us to exclude the former: 
Photons are produced as secondaries in proton-proton (pp) interactions
in the radio jet, they can leave the source freely, and subsequent
interactions with the extragalactic background light (EBL) result in a
very flat TeV gamma-ray spectrum that is inconsistent with HESS data
(Kachelrie{\ss} et al. 2009b). Moreover, the HESS data are
consistent with a point source at the center of Cen\,A, although the
relatively large angular resolution does not allow one to
distinguish between the core and the inner jet as the TeV source. 
Note that the low flux of TeV gamma-rays observed from Cen A 
does not allow to narrow down the source region studying the
time-variability, as e.g.\ in the case M\,87, where the strong
time-variability observed~(Acciari et al. 2009)
excludes the outer jet region as the source of TeV photons,
challenging even the knot HST-1 as the source. Moreover, the large
Lorentz factors implied by the simplest self-synchrotron Compton
models are often not supported by radio data (Piner et al. 2008).
While hadronic models for the origin of TeV gamma-rays have their own 
problems, they offer an alternative source of TeV photons that may avoid 
partly the problems of leptonic models. 

Hadronic models proposed for Cen~A or similar radio galaxies
come in a big variety: There 
are models where TeV gamma-rays are produced in the immediate
vicinity of the supermassive rotating Kerr black hole [Neronov 
\& Aharonian (2007), Rieger \& Aharonian (2008)], in the inner jet
via proton-synchrotron model of Reimer et al. (2004) and
Orellana and Romero (2009), or in
the large-scale jet via proton-proton interactions [Hardcastle et al. (2008)]. 
Only the latter one is disfavoured by the HESS data.

We shall not consider hadronic models using the jet as acceleration site 
(e.g.\  Reimer et al. (2004),  Orellana and Romero (2009), 
Hardcastle et al. (2008)), but shall concentrate here on the second model 
discussed by Kachelrie{\ss} et al. (2009a): This model is based on 
hadronic acceleration near the AGN core, and has the virtue of predicting 
TeV gamma-ray fluxes that are in good agreement with the later HESS 
observations. Moreover, this kind of model is in a sense complementary
to leptonic models for acceleration close to the core: While in the latter 
models TeV photons can escape only for a low photon background in the
central AGN region (Rieger \& Aharonian, 2008 and 2009), hadronic models
require large UV and/or IR backgrounds for the efficient production
of UHECR secondaries. 

The aim of this article is twofold. First, we want to illustrate why 
TeV gamma-rays can escape in hadronic models from an AGN core, although the 
optical depth $\tau_{\gamma\gamma}$ is extremely large, 
$\tau_{\gamma\gamma}(E\sim{\rm TeV})\sim 10^3$. To be more general,
we consider various modifications of our original model, examining e.g.\ the
influence of an additional IR background and a variation of the maximal 
energy $E_{\max}$ of the accelerated protons. Applying our results to the 
special case of Cen\,A is our second aim. In particular, we discuss 
if and how the various observations  of Cen\,A
from the far infra-red to the UHE range can be reconciled.

The outline of the paper is as follows. In Sec.~2, we recall our model in
its original version that is characterized by the dominance of UV photons
in $p\gamma$ and $\gamma\gamma$ interactions. In Sec.~3, we extend these
calculations, varying the photon background density in the core region
as well as $E_{\max}$. We summarize in Sec.~4.

\begin{figure}[h]
\begin{center}
\includegraphics[width=0.95\columnwidth,height=0.8\columnwidth]{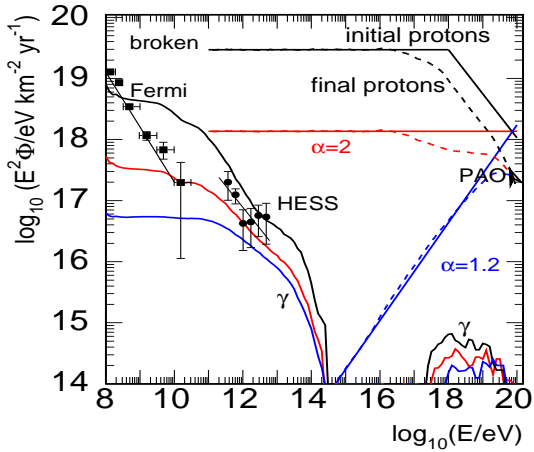}
\caption{\small
Photon (solid lines) and proton (dashed) fluxes from Cen\,A, 
normalized to the PAO results, for different proton injection spectra (solid):
broken power-law (black), power-law with $\alpha=2$ (red), and  
power-law with $\alpha=1.2$ (blue), compared to the data of HESS 
and Fermi-LAT.
\label{core0-fig}}
\end{center}
\end{figure}

%%%%%%%%%%%%%%%%%%%%%%%%%%%%%%%%%%%%%%%%%%%%%%%%%%%%%%%%%%%%%%%%%%%%%%%%%%%
\section{UV dominated background}
We start by recalling the general features of our model; for more details we
refer to Kachelrie{\ss} et al.~2009a.
Acceleration near the core of the AGN can proceed via shock acceleration in 
accretion shocks or via acceleration  in regular fields. 
We do not describe the acceleration process microscopically but rather assume 
that acceleration to ultra-high energies (UHE) takes place in the AGN core and 
produces a power-law flux of UHECRs, $dN/dE\propto E^{-\alpha}$.
We performed calculations for different choices of $\alpha$ and for a
broken power-law spectrum, with  $\alpha _1=2.6$ for $E>10^{18}$\,eV and 
$\alpha _2=2$ for $E<10^{18}$\,eV. In order to account
for the PAO observations we set in this section $E_{\rm max}=10^{20}$~eV.

The spectral energy distribution (SED) of AGNs shows partly a ``big
blue bump'' that is thought to be thermal emission in the UV range
from an accretion disk.  Because of the obscuring dust lane, the UV
emission of Cen\,A cannot be observed directly, and we used therefore
as theoretical model the emission from an optically thick,
geometrically thin accretion disk (Shakura and Syunyaev 1973).  More
specifically, we used $M=1\times 10^8M_\odot$ for the mass of the
black hole, $\dot M=6\times 10^{-4}M_\odot/$yr for the accretion rate
(Evans et al. 2004) and chose the radius of the last stable orbit for
a Schwarzschild black hole $R_0=3R_s=9\times 10^{13}$\,cm as the
smallest radius of the acceleration and emission
region. Since the surface brightness drops fast with
  the radius, most of the radiation is emitted close to the core. To
simplify the treatment, we described the emission region as a sphere
of radius $R_1=15R_s$ filled with a homogeneous, isotropic photon
field of constant density $n_{\rm UV} (\varepsilon)$, as
  discussed in more detail in Kachelrie{\ss} et al.~2009a.  In
addition, we assume that a hot corona produces an X-ray component with
$n_{\rm X} (\varepsilon)\sim \varepsilon ^{-1.7}$. The photon
densities $n_i(\varepsilon)$ are normalized from the relation
\begin{equation} 
  L_i=\pi R_1^2 c\int \!d\varepsilon\;
  \varepsilon\,n_i(\varepsilon)=\eta_i\,\dot Mc^2\,,
\label{eq:luminosity}
\end{equation} 
using the measured luminosity $L_X=4.8\cdot 10^{41}$ erg/s in the 2-10 keV range for X-ray photons
(Evans 2004 and Markowitz 2007) and choosing the efficiency $\eta _{\rm UV}=10$\% in the UV range.
Finally, we assumed that  IR photons are distributed on a much larger spatial scale such 
that the corresponding density $n_{\rm IR} (\varepsilon)$ can be neglected.

\begin{figure}[h]
\begin{center}
\includegraphics[width=0.95\columnwidth,height=0.8\columnwidth]{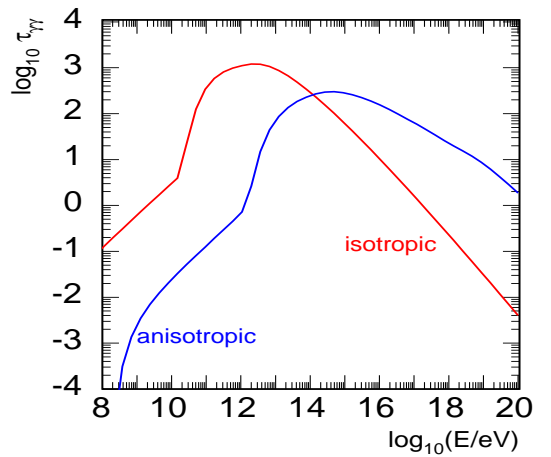} 
\caption{\small
The optical depth $\tau_{\gamma\gamma}$ for photon-photon interactions inside 
the UV emission region ($r<R_1$, red line) and outside in the 
quasi-collinear regime ($r>4R_1$, blue line) only on UV photons. 
\label{depth0-fig}}
\end{center}
\end{figure}

Photons, electrons and positrons (e/m particles) produced in $p\gamma$  
interactions further cascade on the UV photon background via  
pair production, inverse Compton scattering, and synchrotron emission. 
The latter contribution to the cascade, due to the regular and turbulent 
magnetic fields, requires a separate discussion:
Electrons move along the regular field lines, emitting synchrotron 
and curvature radiation. We describe these processes by an effective random 
magnetic field. Because of the rather high energies of the electrons 
involved, synchrotron emission proceeds in the quantum regime, i.e.\ with
$\chi\equiv eB_{\perp}E_e/m_e^3\gsim 1$: Consequently, electrons emit a small 
number of high-energy photons instead of producing a semi-continuous spectrum
of low energy photons.
Thus, the cascade develops in a very peculiar way: Photons are 
pair-producing on the background 
photons, while electrons contribute mainly to the cascade via synchrotron radiation. We verified that the 
resulting spectrum depends only weakly on the assumed strength of the effective magnetic field
strength $B$: Changing $B$ by four orders of magnitude from $10^{-2}$ up to 
$10^2$\,G has only a minor impact on final photon spectra leaving the 
source\footnote{For high-energy electrons, the characteristic 
parameter for the synchrotron emission $\chi$ remains
large over a wide range of magnetic field strengths, which thus assures that
the emission proceeds in the quantum regime.}.

The predicted photon spectra after cascading in the  EBL are compared 
for three different choices of injection proton spectra 
to recent HESS and preliminary Fermi-LAT data in Fig.~\ref{core0-fig}. 
Remarkably, the obtained spectral
shape in the TeV range depends only weakly on the proton spectral 
slope  and agrees well with the 
HESS data for all the three cases considered. 
However, the broken power-law proton spectrum results in a too high photon flux
in the GeV range,  in contradiction to the Fermi data.

Let us now discuss the feature of our numerical results that may be most 
interesting from a general point of view: How can TeV gamma-rays leave
the AGN core despite the large UV density?  To illustrate this problem, 
we plot in Fig.~\ref{depth0-fig}  as a red line the optical depth
  $\tau _{\gamma \gamma}(E)$ as a function of energy for a photon 
  propagating through the distance range
$[R_0,R_1]$ corresponding to the UV emission 'photosphere'.
 The optical depth  is of order $10^3$ in the TeV
range and one expects that most electromagnetic energy cascades down 
in the 10\,GeV range, forming a characteristic spectral plateau. 

There are two different contributions to the final TeV photon flux.
To understand the first one, it is useful to consider separately
the development of the e/m cascade inside ($R< R_1$) and outside 
($R> R_1$) of the emission region. In the former case, the angular 
distribution of UV photons is isotropic. The emission photosphere is
opaque for TeV gamma-rays. However, UHE photons ($E\gsim 10^{18}$ eV) 
interact with the background photon fields in the deep Klein-Nishina 
regime ($E\,\varepsilon\gg s_{\rm thr}=4m_{e}^2$). Since the pair production 
cross section decreases as $1/E$ above the threshold, the core region 
appears to be transparent for them. 
Outside the emission region UV photons are anisotropic. At large distances 
$r\gg R_1$ their density goes down
as $n_{\rm UV}\sim (R_1/r)^2$ and their momentum vectors obey 
\begin{equation} 
% \cos \theta = \frac {\vec p \vec r}{pr}\geq 1-(R_1/r)^2\, .
 \cos \theta = \frac {{\bf p \cdot r}}{{\bf |p||r|}}\geq 1-(R_1/r)^2\, .
\end{equation} 
Thus ${\bf p}$ becomes more and more collinear with the 
momentum vector of the outgoing e/m cascade particles.
At large enough distances, UHE photons interact therefore with the UV 
background close to the threshold, 
$s=2E\,\varepsilon\,(1-\cos\theta)\sim 4m_e^2$, and the produced TeV photons 
leave the source vicinity freely. To illustrate this picture, we show  in 
Fig.~\ref{depth0-fig} also the optical depth for a photon injected at 
$r=4R_1$  as a blue line. 

The second contribution to the final TeV photon flux are muons created
in $p\gamma\to n\pi^+$ reactions. Their decay length becomes for 
$E\approx 10^{18}$\,eV comparable to the size $R_1$ of the isotropic UV region. 
Thus most UHE muons decay outside in the ``collinear region,'' where the 
cascade stops above the TeV scale.

In summary, the TeV gamma-ray flux is formed by decays of UHE muons
and by re-shuffling UHE photons with $E\gsim 10^{17}$\,eV in 
%interactions with 
the anisotropic UV background surrounding the source region.
 
It is worth stressing the salient points of this picture: First, both
processes require a sufficiently high flux of UHE photons or muons.
Such particles can only be generated as secondaries from accelerated
hadrons.  Second, the observed gamma-ray spectrum is essentially
formed in the source vicinity. Subsequent interactions with the EBL
lead only to minor modifications. Third, the e/m cascade in the
  source vicinity, which re-shuffles photons of energies $E\gsim
  10^{18}$ eV  to the TeV
  range, includes a large number of particle generations (emission
  steps). As a consequence, the photon spectral shape in the TeV range
  depends rather weakly on the proton injection spectrum; the latter
  rather defines the normalization of the spectral 'plateau'
    in the GeV range.

By contrast, photons leave the source freely when they are produced
by protons accelerated over long distances ($\sim$\,kpc) in the radio jet.
The further re-shuffling of UHE photons in interactions with the EBL
results in a very flat TeV gamma-ray spectrum (Kachelrie{\ss} et al.~2009a), 
in contradiction to the HESS observations.

%%%%%%%%%%%%%%%%%%%%%%%%%%%%%%%%%%%%%%%%%%%%%%%%%%%%%%%%%%%%%%%%%%%%%%%%%%%%%%%%%%%%%%%
\section{Varying the photon background and $E_{\max}$}
\begin{figure}[h]
\begin{center}
\includegraphics[width=0.95\columnwidth,height=0.8\columnwidth]{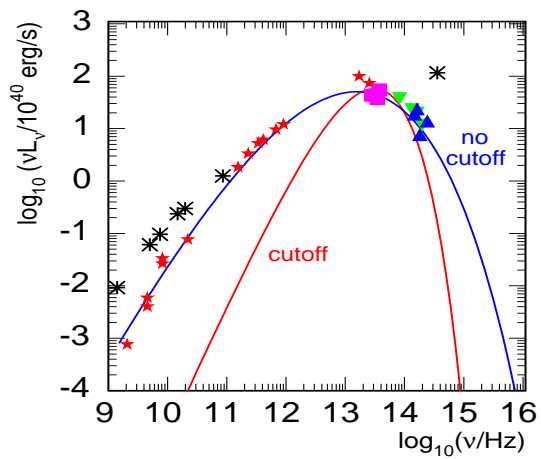}
\caption{\small 
Parametrizations of the IR part of the SED towards
smaller frequencies, i) cutoff (red), and 
ii) no cutoff (blue) in the far IR, together with observational data 
points.
\label{sed-fig}}
\end{center}
\end{figure}

Several of the assumptions underlying our analysis in the last section 
were chosen on the basis of weak empirical evidence. In detail, these 
assumptions are:
\begin{enumerate}
\item A negligible impact of the IR photons on the cascade
  development. 
\item The large value of $E_{\max}=10^{20}$~eV. 
\item The size of the UV bump, with an efficiency of
  $\eta_{\rm UV}=10\%$, as defined in Eq.~(\ref{eq:luminosity}). 
\end{enumerate}

The objective of this section is to discuss how the results
obtained 
in Kachelrie{\ss} et al.~2009a, and summarized in Fig.~\ref{core0-fig}
are modified by relaxing the previous assumptions, i.e.
varying the background photon field of the source
 as well as $E_{\rm max}$.  
In the following, we shall use the present estimate $M=5\times 10^7 M_\odot$ 
for the black hole mass of  Centaurus~A (Cappellari et al. 2008),
which corresponds to the inner radius of the acceleration/emission  region
$R_0\simeq 4.5\times 10^{13}$ cm. 
%Our default choice for the AGN efficiency in the  UV range is
%$\eta _{\rm UV}=5$\%, which corresponds to the UV luminosity 
%$L_{\rm UV}=1.8\times 10^{42}$ erg/s,
%and  we use $E_{\max}=10^{20}$ eV as the maximal UHECR energy.
We shall use always a $dN/dE\sim E^{-2}$ spectrum for the injected protons,
and if not otherwise stated $\eta _{\rm UV}=5$\%.

\subsection{Infrared background}

We start by investigating the effect of varying the IR
background on the TeV photon spectra. 
Conventionally, the dominant
contribution to the SED in the IR range is supposed to be formed by
the thermal emission from the dust torus and is thus distributed on
multi-parsec scales.  Nevertheless, recent interferometry measurements
by Meisenheimer et al.  (2007) indicate that a significant part of the
near IR emission (from 60\% at $\lambda=13\,\mu$m up to 80\% at
$\lambda=8\,\mu$m) is due to a compact unresolved source of size
$R_{\rm IR}<0.2$\,pc.  (Note, however, that those results are
challenged by Radomski et al. 2008.) Moreover, it has been speculated
by Meisenheimer et al. (2007) that the whole SED from the IR down to
the radio range is mainly generated by a tiny synchrotron source of
the size $R_{\rm IR}\sim 0.01$\,pc. Since the optical depth scales as
$\tau\propto R^{-1}$, interactions with IR photons close to the core
should seriously modify the picture discussed in Sec.~2.
%%%

In order to analyze the phenomenological consequences
  of an IR background on the TeV photon spectra, we consider two
  different parametrizations of the SED shown in Fig.~\ref{sed-fig}:
  i) With a cutoff in the far IR, as one expects if IR photons
  are due to thermal emission of the torus (Marconi et al.~2000); ii) the whole
  SED from the IR down to the radio range emitted by the same source
  of size $R_{\rm IR}$ as suggested by Meisenheimer et al. (2007).
Additionally, we have varied the spatial size of the IR source
over two orders of magnitude, $R_{\rm IR}=$1, 0.1, and 0.01\,pc.

\begin{figure}%[h]
\begin{center}
\includegraphics[width=0.95\columnwidth,height=0.8\columnwidth]{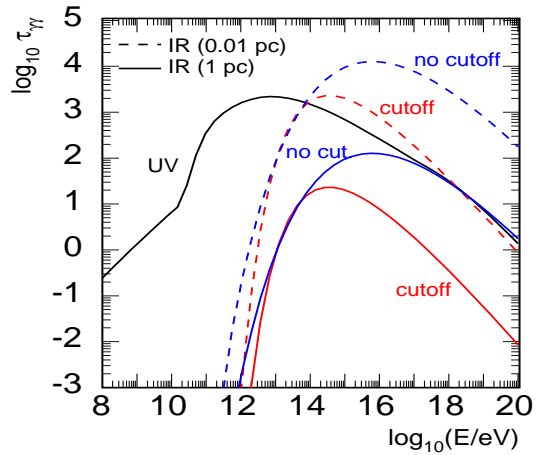}
\caption{\small Photon optical depth due to interactions with the UV
  (solid black line) and IR backgrounds for different
  parametrizations of the infra-red part of the SED:
  cutoff (red), and no cutoff (blue); for $R_{\rm
    IR}=0.01$ (dashed) and 1\,pc (solid).
\label{depth-fig}}
\end{center}
\end{figure}

Before showing the results let us first discuss how
  the presence of the IR background modifies the interaction depth
  of photons and protons.
The relative size of the contribution of the (isotropic and anisotropic) 
UV  and
the IR backgrounds to the total
optical depth of photons is illustrated for
the two parametrizations of the SED and two source sizes, $R_{\rm IR}=1$ and
0.01\,pc in Fig.~\ref{depth-fig}.  \ignore{ For two options, a small
  source with $R=0.01$\,pc with either a soft or no cut-off at all,
  $\tau_{\gamma \gamma}^{\rm IR}$ is comparable or even larger than
  $\tau_{\gamma \gamma}^{\rm UV}$ above 100\,TeV.  In particular, the
  core region becomes now also opaque for UHE photons, $E\gsim
  10^{17}$\,eV, which can interact with the dense IR background.  Note
  that this on one side undermines the reshuffling mechanism described
  in Sec.~2, but on the other side means that the exact value of the
  UV luminosity becomes less important for $\gamma\gamma$ (and
  $p\gamma$ above threshold) interactions.  } 
Let us first consider the contribution to the optical depth 
due to photon interactions with the IR background in the UV emission region,
i.e.~at $R<R_1\approx  10^{-4}$\,pc, which can be obtained 
by rescaling the different curves of Fig.~\ref{depth-fig}:
$\tau^{\rm IR}_{\gamma\gamma}(R_1)=
\tau^{\rm IR}_{\gamma\gamma}(R_{\rm IR})\,R_1/R_{\rm IR}$.
 For most of the cases considered,  photon interactions with   IR photons
  at $R<R_1$ are a small correction compared to interactions with the
  UV background. Thus, the situation is analogous
  to the one discussed previously:
Photons created in the UV region with energies
above $10^{18}$~eV can escape. Once they enter the IR region they
cascade down to the 10\,TeV energy range, now both due to 
 interactions with quasi-collinear UV photons and with the
IR photons; see
  Fig.~\ref{depth0-fig}. Therefore the reshuffling mechanism still
works.  This is true except for the case of an extreme compact IR
background that explains the full SED, i.e.\ $R_{\rm
  IR}=0.01$\,pc without cutoff. In this case,
interactions with IR photons dominate even inside $R_1$ and make the
UV region almost opaque for photons with energies below $E\approx
10^{20}$~eV.

\begin{figure}
\begin{center}
\includegraphics[width=0.95\columnwidth,height=0.8\columnwidth,angle=0]{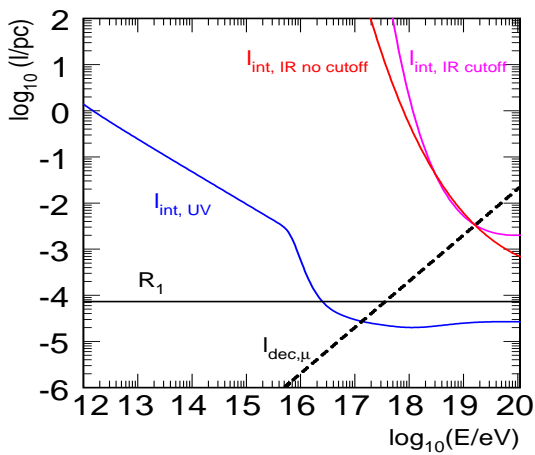}
\caption{\small Comparison of the size of the UV source $R_1$ (black
  solid line) with the interaction length for $p\gamma$ scattering on
  UV (blue solid) and IR photons with  cutoff (solid magenta) and
  no cutoff (solid red), together with the muon decay length (black
  dashed line), for $\eta_{\rm UV}=5\%$ and $R_{\rm IR}=0.01$\,pc.
\label{pg-lengths}}
\end{center}
\end{figure}

As far as the proton interactions with the UV and IR
  background are concerned, we show in
Fig.~\ref{pg-lengths}
 the interaction lengths for $p\gamma$ pion
interactions assuming $R_{\rm IR}=0.01$\,pc  compared to the size 
of the UV region $R_1$ and the muon decay length. The threshold energy for 
$p\gamma_{\rm UV}$ and $p\gamma_{\rm IR}$ is roughly $10^{16}$\,eV and 
$10^{18}$\,eV, respectively. For this choice of parameters, the interaction 
depth $\tau_{p\gamma}$ above the threshold is of  order a few for both photon 
backgrounds, while rescaling e.g.\ to $R_{IR}=1$~pc gives 
$\tau_{p\gamma}\propto R^{-1}\sim$\,few~\% in the IR region. 
Figure~\ref{pg-lengths} shows also that muons with
energies above a few$\times 10^{17}$\,eV escape from the UV region 
before decaying.

\begin{figure*}[t]
\begin{center}
\begin{tabular}{cc}
\includegraphics[width=0.95\columnwidth,height=0.8\columnwidth]{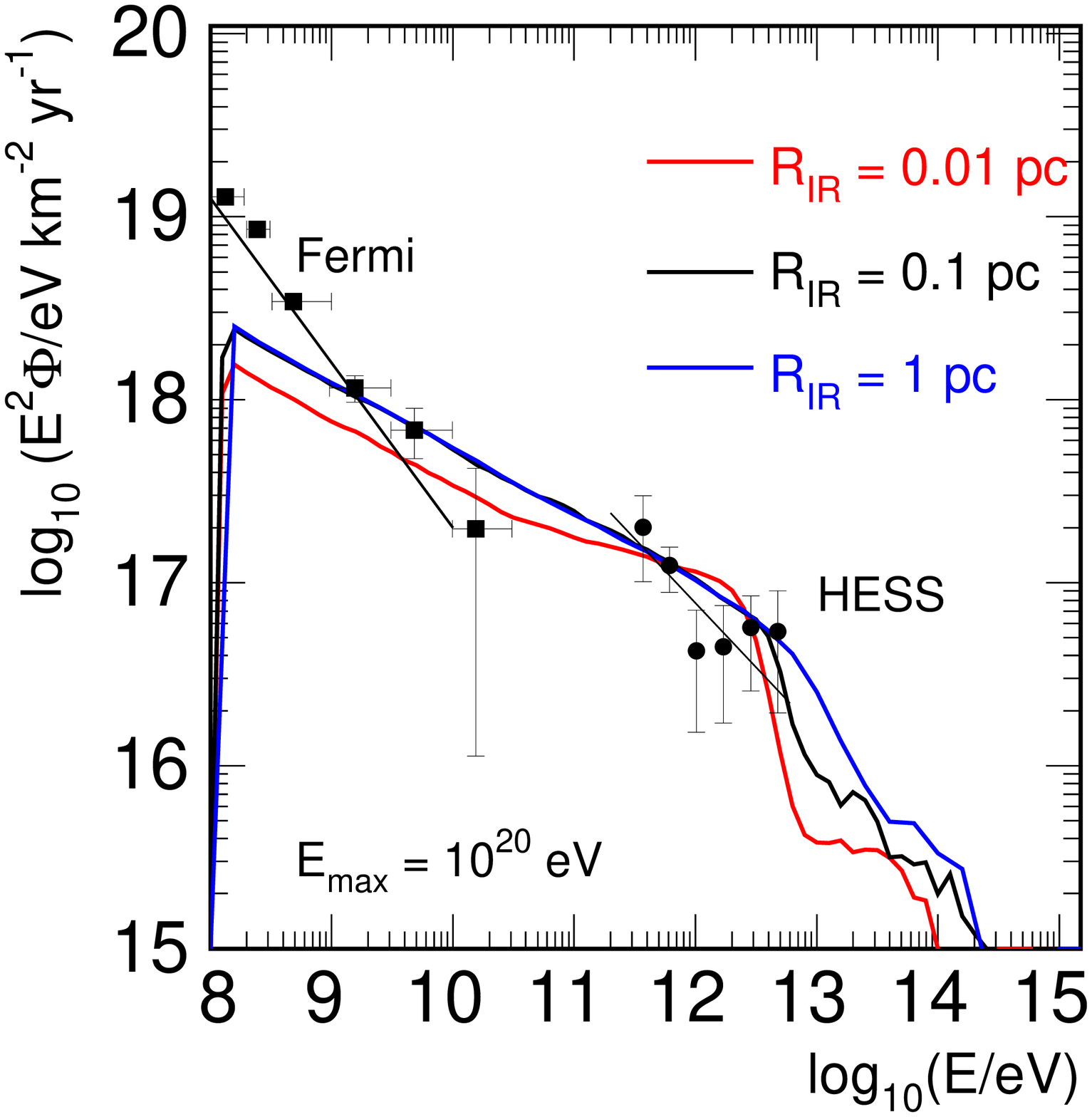}
&
\includegraphics[width=0.95\columnwidth,height=0.8\columnwidth]{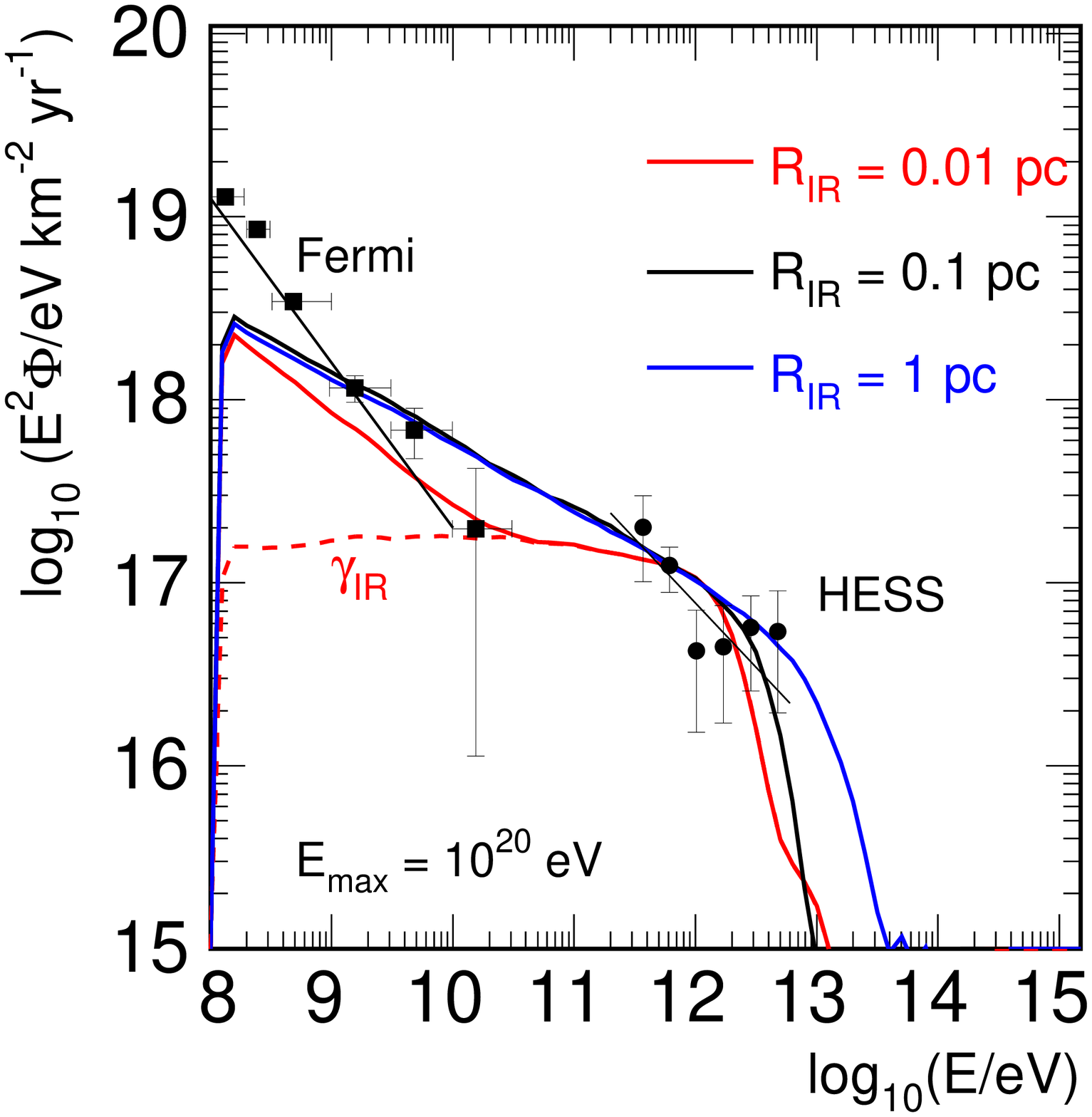}
\end{tabular}
\caption{\small
Photon  fluxes from Cen\,A normalized to the HESS observations
for different sizes of the IR source and for different parametrizations 
of the SED: cutoff in the far IR (left), no cutoff in the far IR 
(right); $E_{\max}=10^{20}$\,eV.}\label{spec20-fig}
\end{center}
\end{figure*}

We now discuss our new results for the photon spectrum
  from Cen\,A and, in particular, if our numerical results agree with
  our expectations from the $\tau_{\gamma\gamma}$ and $l^{\rm
    int}_{p\gamma}$ plots. In all cases we normalize the spectra to
  the HESS results.
In Fig.~\ref{spec20-fig}, photon fluxes calculated for the maximal
UHECR energy $E_{\max}=10^{20}$ eV are plotted for three different
sizes of the IR source, with a cutoff in the left and no cutoff in the
right panel. Compared to case of only UV background shown in Fig.~1,
the cutoff observed in the gamma ray flux is more
pronounced and shifted to lower energies. Still the shape is
consistent with the HESS data, except perhaps for the most drastic
case, $R_{\rm IR}=0.01$\,pc without a cutoff.  For all the cases
considered, the obtained spectra do not contradict the Fermi data in
the GeV range.

How could we obtain a substantial TeV photon flux, although the core 
region is completely opaque for UHE photons? There are again two effects
that explain the TeV photon flux:
First, the probability for
$p\gamma$ interactions on IR photons is non-negligible, and for some
parameters even comparable to the one on UV, cf. Fig.~\ref{pg-lengths}. 
Since $\tau_{p\gamma}\sim\max\{\tau_{\gamma\gamma}\}/2000$, the distribution
of the starting points for these cascades is shifted outwards and
the remaining path length through the photon gas thus reduced.
Second, the muon mechanism already encountered in the ``UV only'' case is 
again operative.
The relative importance of these two effects depends on the chosen 
parameters, in particular the size of the IR and UV regions,  
the shape of the IR cutoff as well as $E_{\max}$.

In order to understand this better, we show in the right panel of 
Fig.~\ref{spec20-fig} for the case $R_{\rm IR}=0.01$\,pc not only the 
total photon flux, but separately by a dashed red line those photons 
that were generated by electromagnetic cascades initiated outside the
UV emission  region, i.e.~at $R>R_1$. Note that this includes e/m cascades
 initiated by
decaying muons as well as protons interacting at $R>R_1$. 
Clearly, the flux above 1\,TeV is dominated by photons that  are 
created as final products of e/m cascades initiated outside the
 UV photosphere.

It is remarkable how similar the results presented in Fig.~\ref{spec20-fig}
are, 
being practically independent on the spectral shape and
the spatial extension of the IR background. The amount of TeV photons 
produced depends on the flux of photons  generated in the UV region at 
energies above $10^{18}$\,eV and being reshuffled in the source 
vicinity (at $R>R_1$),
 of photons originating from proton-gamma interactions at $R>R_1$, which come
 mainly from  the decays of produced neutral pions, and of 
high energy muons created in the UV emission region and decaying outside it. 
Whereas the latter contribution is only determined by $\eta_{\rm UV}$ 
and $R_1$, the two former ones explicitly depend on the properties of the 
IR background.
However, e/m cascades stop  around 10\,TeV for all the IR backgrounds 
considered, cf.~Fig.~\ref{depth-fig}. This value corresponds roughly to
the cutoff visible in  Fig.~\ref{spec20-fig}.

\subsection{Maximum energy}

%%%
Deriving the previous results a maximum energy of $10^{20}$~eV
  for the protons was assumed. Acceleration of protons to these energies 
requires magnetic field strengths of order
$B\sim$\,kG.  This is an order of magnitude higher than what one would
expect from equipartition arguments, $B^2/(8\pi)\sim L/(4\pi R_1^2c)$,
and two orders below the estimate developed by Bicknell \& Li (2007)
using a Poynting flux dominated jet.
An enhancement of the field strength may, in principle, be expected
due to the back-reaction of accelerated particles (Bell \& Lucek
2001). However, alternative pictures where either the core region acts
as a pre-accelerator, producing protons up to $E_{\max}\sim 10^{18}$
eV, while acceleration to ultrahigh energies takes place in the lobes
(Hardcastle et al.~2008) or Cen\,A produces no UHECRs at all are
certainly also possible.

\begin{figure*}[!htb]
\begin{center}
\begin{tabular}{cc}
\includegraphics[width=0.95\columnwidth,height=0.8\columnwidth]{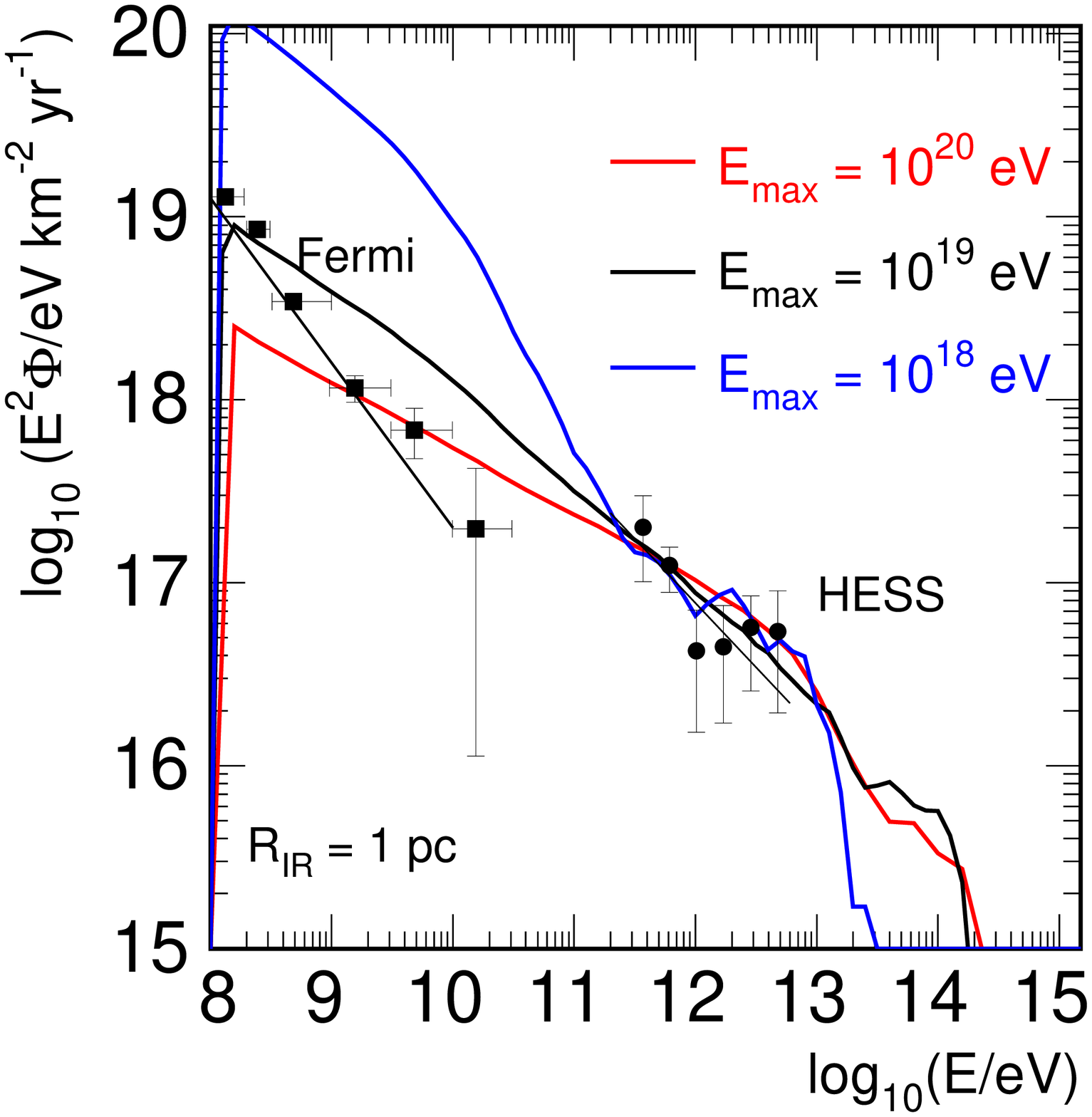}
&
\includegraphics[width=0.95\columnwidth,height=0.8\columnwidth]{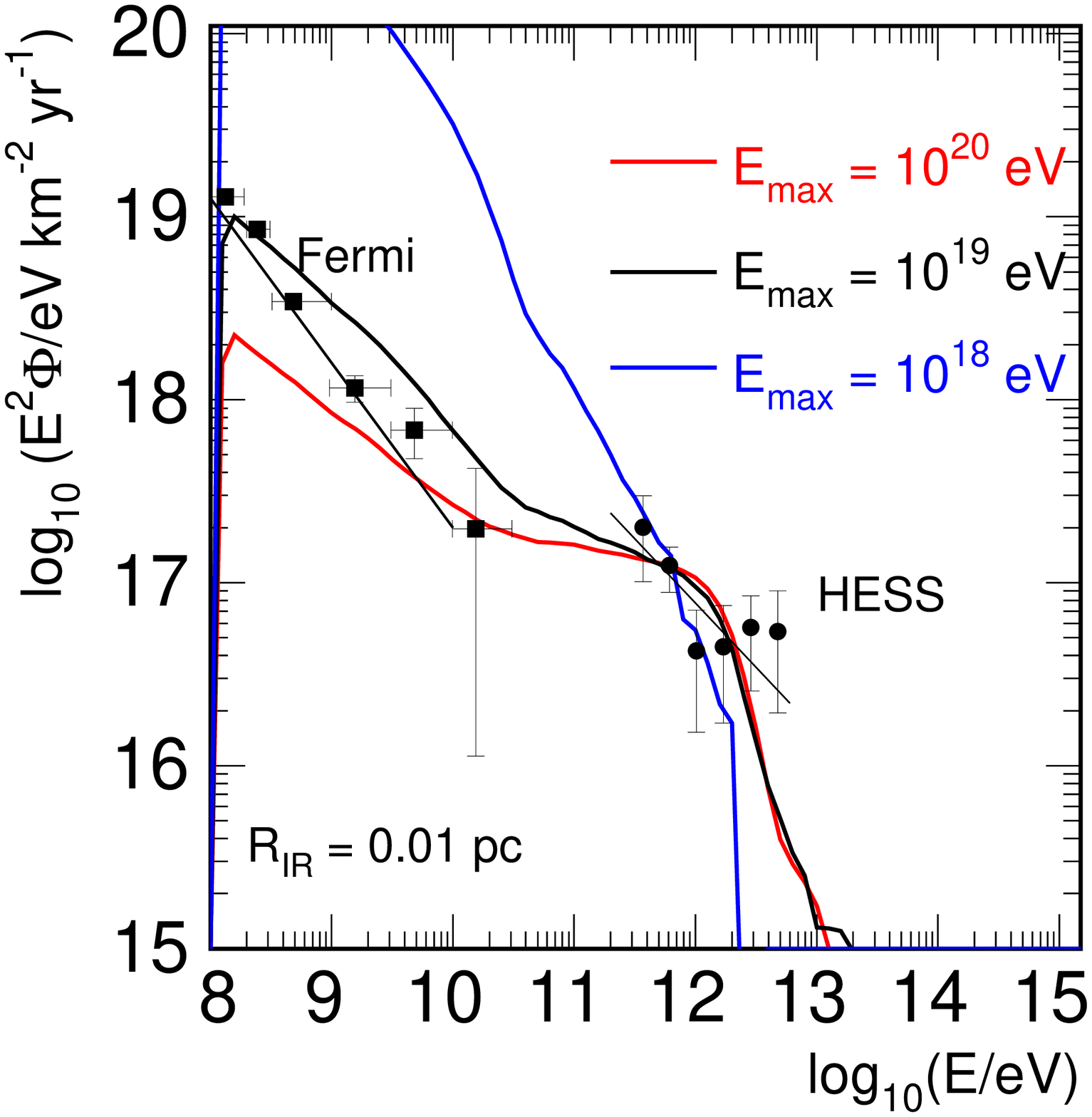}
\end{tabular}
\caption{\small
Photon  fluxes from Cen\,A calculated for different $E_{\max}$ and for 
a cutoff in the far IR and $R_{\rm IR}=1$ pc (left), 
without cutoff and $R_{\rm IR}=0.01$ pc (right).
\label{78mixed}}
\end{center}
\end{figure*}

In this subsection we analyze the consequences of lowering the maximal energy. 
In Fig.~\ref{78mixed},
 we show the photon fluxes for three different values
of the maximal UHECR energy, $E_{\max}=10^{18}$, $10^{19}$ and $10^{20}$\,eV,
and for two different scenarios for the IR background. Reducing  $E_{\max}$
leads to a smaller number of UHECR above the threshold for interactions on 
the IR and a smaller number of muons decaying outside, reducing in turn 
the TeV photon flux. Keeping thus the TeV flux fixed, the  GeV flux 
increases, exceeding for $E_{\max}=10^{18}$\,eV and marginally  
for $E_{\max}=10^{19}$\,eV the Fermi data.

\subsection{Ultraviolet background}

The presence of a blue/UV bump in the SED spectrum is
a general characteristic of AGNs.
In the analysis reported in Kachelrie{\ss} et al.~2009a, the 
luminosity of the UV bump was set to 10\% of the accretion
rate. Since
no observational data for Cen\,A exist in this range, the UV emission could be
considerably smaller than we assumed. This is also suggested by the
low accretion rate of Cen\,A, that indicates that the accretion is
advection-dominated.

\begin{figure*}[!htb]
\begin{center}
\begin{tabular}{cc}
\includegraphics[width=0.95\columnwidth,height=0.8\columnwidth]{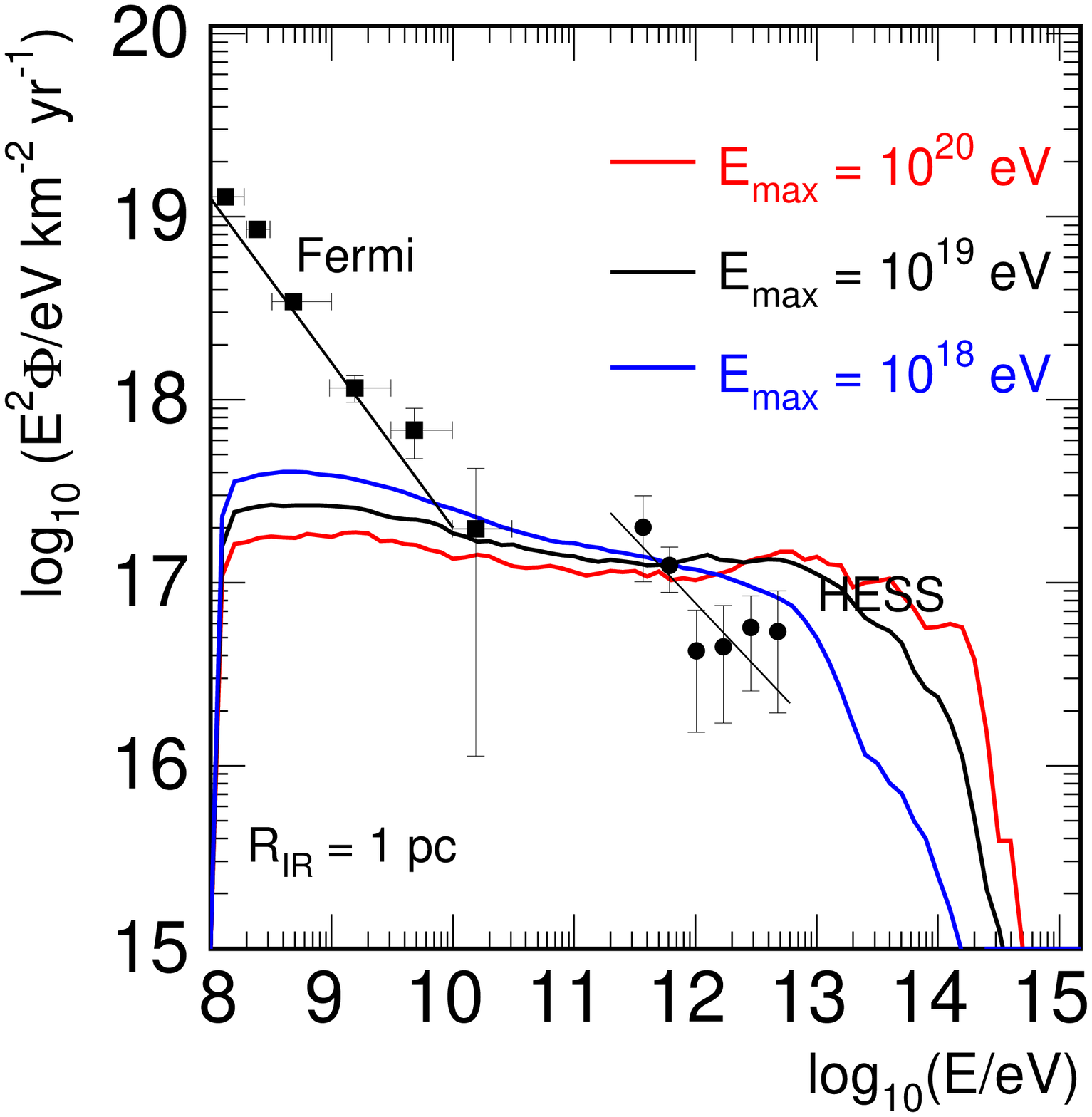}
&
\includegraphics[width=0.95\columnwidth,height=0.8\columnwidth]{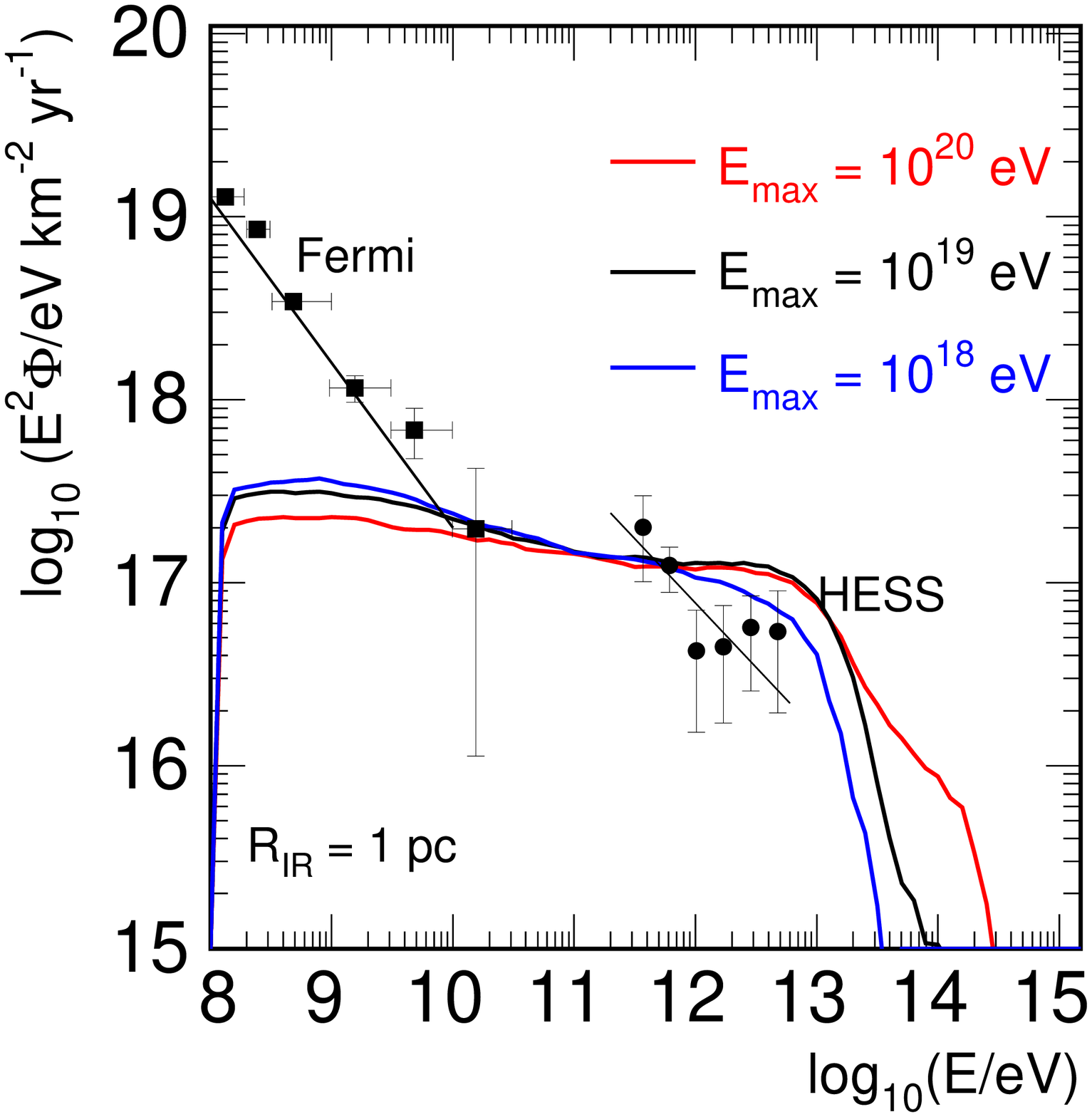}
\end{tabular}
\caption{\small
Photon  fluxes from Cen\,A for $R_{\rm IR}=1$\,pc and $\eta_{\rm UV}=0.1\%$;
cutoff in the far IR (left) and without the cutoff in the far IR 
(right).
\label{fig10}}
\end{center}
\end{figure*}

Therefore  we analyze in the following the consequences of modifying the UV
background. Reducing the relative importance of the UV
background should lead to a larger fraction of $p\gamma$ interactions 
on the IR, thereby enhancing the TeV and reducing the GeV flux.
This effect is demonstrated in
Fig.~\ref{fig10}, 
where the photon fluxes for $\eta_{\rm UV}=0.1\%$ are shown for two IR
backgrounds and $R_{\rm IR}=1$~pc. The GeV photon flux is now as
expected suppressed and even for $E_{\max}=10^{18}$\,eV  more or
  less consistent with Fermi data, apart from a flatter than observed
  spectral shape. This shape coincides with the flat
  contribution from the $R>R_1$ expected in the case with important UV
  component; see dashed lines in the right panel of Fig.~\ref{spec20-fig}. 
Both with and without cutoffs, one would expect
  from the criterion $\tau_{\gamma\gamma}=1$ that no photons with
$E\gsim $few TeV should be observed from Cen\,A, while the actual flux
for these parameters extends up to few hundreds of TeV. The extension
to high energies is most prominent for a cutoff in the IR and large
$E_{\max}$.  Both conditions lead to a larger fraction of UHE protons
with energy above the pion production threshold interacting with
  the IR photons, and thus to the generation of electromagnetic
cascades at relatively large distances from the core.

However, there is a price to be paid
 for the suppression of the GeV photon
flux: Since $\tau_{p\gamma}$ is now only of order
5\% above the threshold, a larger CR flux is required to fit the HESS data.
In particular, the CR luminosity required
for the parameters used in the left panel of Fig.~\ref{fig10} 
%\ref{78mixed}
 is  $L_{\rm CR, tot}=2\times 10^{43}$erg/s using $E_{\min}=10^{9}$\,eV. 
This value should be compared to
the total bolometric luminosity of Cen\,A, $L_{\rm bol}=10^{44}$\,erg/s.
Hence such a scenario requires an uncomfortably large fraction of the total
energy output of the AGN core channeled into CRs.

This problem can be ameliorated reducing the size of the IR source. 
Choosing $R_{\rm IR}=0.01$\,pc, the interaction depth $\tau_{p\gamma}$ 
in the IR region increases to $\tau_{p\gamma}\sim$\,few above the threshold
and as a result the CR luminosity is reduced. 
In the particular case of an IR background without cutoff
in the far IR, the required 
CR luminosity is a factor five smaller, $L_{\rm CR, tot}=4\times 10^{42}$\,erg/s 
for $E_{\min}=10^{9}$\,eV, but the cutoff is in the 10\,TeV range. 
Thus we predict an extension of the VHE photon flux from Cen\,A into the 
100\,TeV range, only if the UV background in Cen\,A is sufficiently
large, $\eta_{\rm UV}\gsim 1\%$.
% or the source of the IR background is
%extremely compact, $R_{\rm IR}\sim 0.01$\,pc.

%%%%%%%%%%%%%%%%%%%%%%%%%%%%%%%%%%%%%%%%%%%%%%%%%%%%%%%%%%%%%%%%%%%%%%%%%%%
\section{Summary}

We have investigated the hypothesis that the TeV photons observed by HESS 
from Cen\,A are produced as secondaries in hadronic interactions.
Varying the parameters of the photon background in the IR and UV, we
have identified cases for which the predicted VHE photon spectrum
extends beyond the energy cutoff predicted naively from the
condition $\tau_{\gamma\gamma}(E_{\rm cut})=1$. Such cases can be as
diverse as i) a dominant UV background with a sufficiently diffuse 
IR background or ii) a low UV background with a compact IR background
on a (sub-) pc scale. Applied to Cen\,A, both cases predict photon spectra
that are compatible with HESS and Fermi data, although for the second 
option the obtained spectral shape in the TeV range is significantly flatter
 than observed experimentally.  Additionally, the first
case is also consistent with the possibility of Cen\,A as an Auger
source. Within our scenario, the cutoff in the VHE photon flux from 
Cen\,A could be in the $10\div 100$\,TeV range, if i) $E_{\max}\gsim 10^{19}$\,eV
and ii) either the UV background in Cen\,A is sufficiently
large, $\eta_{\rm UV}\gsim 1\%$, or the source of the IR background is
extremely compact, $R_{\rm IR}\sim 0.01$\,pc.

Depending on the parameters, different mechanisms contribute to the
photon flux in the multi-TeV region: UHE muons created in $p\gamma\to\pi^+n$
reactions can carry e/m energy efficiently outwards before they decay.
UHE photons interact in the inner part of the AGN core in the Klein-Nishina
regime, channeling therefore also efficiently e/m energy outwards 
before they interact in the collinear regime. Finally, $p\gamma$
interactions on IR photons enhance additionally the high-energy tail
of the photon flux. All three mechanisms are based on hadronic primaries
and require a flux of UHECRs extending at least to $10^{18}$\,eV.

Viable combinations of IR and UV backgrounds are those that allow
a sufficiently large fraction of UHE photons or muons to traverse 
the UV region and/or that lead to an adequate number of $p\gamma$ 
interactions on the IR, while satisfying luminosity constraints.

Finally, we stress that our findings are not  only valid for Cen\,A, 
but could be applied more generally to other AGN as sources of VHE energy 
photons and in particular to blazars.

%%%%%%%%%%%%%%%%%%%%%%%%%%%%%%%%%%%%%%%%%%%%%%%%%%%%%%%%%%%%%%%%%%%%%%%%%
\section*{Acknowledgments} %If needed

We would like to thank M.J.~Hardcastle for useful discussions.
S.O.\  acknowledges a Marie Curie IEF fellowship from the European Community, 
R.T \  partial support from the Deutsche Forschungsgemeinschaft within the 
SFB 676.


\begin{thebibliography}{}
% References are listed as in the following example, for more examples, please
% see the PASA Style Guide

\bibitem[Abraham, J.(2007) et al.]{}Abraham, J., et al.~(PAO Collaboration)~2007,
Science {\bf 318}, 939.

\bibitem[Acciari, V.~A. (2009) et al.]{}%{Acciari:2009rs}
V.~A.~Acciari {\it et al.}  [VERTIAS, HESS and MAGIC Collaborations],
  %``Radio Imaging of the Very-High-Energy Gamma-Ray Emission Region in the
  %Central Engine of a Radio Galaxy,''
  Science {\bf 325}, 444 (2009).
%  [arXiv:0908.0511 [astro-ph.HE]].
  %%CITATION = SCIEA,325,444;%%


\bibitem[Aharonian, F.(2009) et al.]{}Aharonian, F.,  et al.~(HESS Collaboration)~2009,
Astrophys.\ J.\ Lett.\  {\bf 695}, L40.

\bibitem[Aharonian, F. and Rieger, F (2009)]{}Aharonian, F. and
  Rieger, F.~2009, Astron.\ Astrophys.\ {\bf 506}, 3, L41.


\bibitem[Bell, A.~R., Lucek, S.~G.(2001)]{}Bell, A.~R., Lucek, S.~G.~2001,
 MNRAS {\bf 321}, 438.

\bibitem[Berezinsky, V.(2004) et al.]{}Berezinsky, V.,  
Gazizov, A.~Z.,  Grigorieva, S.~I.~2004,
Nucl.\ Phys.\ Proc.\ Suppl.\  {\bf 136}, 147;
Phys.\ Lett.\ B {\bf 612} (2005) 147.

\bibitem[Bicknell, G.~V.(2007) et al.]{}Bicknell,~G.~V. and Li,~J.,
  %``Linkage between Accretion Disks and Blazars,''
  Astrophys.\ Space Sci.\  {\bf 311}, 275 (2007).
%  [arXiv:0704.2636 [astro-ph]].
  %%CITATION = APSSB,311,275;%%

\bibitem[Cappellari, M.(2008) et al.]{}Cappellari, M., et al.~2008,
 arXiv:0812.1000 [astro-ph]. 

\bibitem[Cheung, T.(2009)]{}Cheung, T., et al.~(Fermi-LAT Collaboration)~2009,
 this volume.

\bibitem[Evans, D.~A.(2004) et al.]{}Evans, D.~A., et al.~2004, 
Astrophys.\ J.\  {\bf 612}, 786.

\bibitem[Hardcastle, M.~J.(2008) et al.]{}Hardcastle, M.~J., Cheung, C.~C.,
Feain, I.~J., Stawarz, L.~2008,
 MNRAS {\bf 24}, 337.

\bibitem[Kachelrie\ss, M.(2009a) et al.]{}Kachelrie\ss, M.,
Ostapchenko, S.,  Tom\`as, R.~2009a,
New J.\ Phys.\  {\bf 11}, 065017.

\bibitem[Kachelrie\ss, M.(2009b) et al.]{}Kachelrie\ss, M.,
Ostapchenko, S.,  Tom\`as, R.~2009b, 
%``Multi-messenger astronomy with Centaurus A,''
  Int.\ J.\ Mod.\ Phys.\  D {\bf 18}, 1591 (2009).
  %%CITATION = IMPAE,D18,1591;%%


\bibitem[Marconi, A.(2000) et al.]{}Marconi, A.,  et al.~2000,
Astrophys.\ J.\  {\bf 528}, 276.

\bibitem[Markowitz, A.(2007) et al.]{}Markowitz, A.,  et al.~2007,
Astrophys.\ J.\  {\bf 665}, 209.

\bibitem[Meisenheimer, K.(2007) et al.]{}Meisenheimer, K., et al.~2007,
 Astron.\ Astrophys.\ {\bf 471}, 453.

\bibitem[Neronov, A (2007) et al.]{}Neronov, A., \& Aharonian, F. A. 2007, 
Astrophys.\ J.\ {\bf 671}, 85.

\bibitem[Orellana (2009)]{}Orellana, M. and Romero, G.~E. 2009,
%``A model for the electromagnetic spectrum of the inner jets of Cen A,''
  AIP Conf.\ Proc.\  {\bf 1123}, 242 (2009).
  %%CITATION = APCPC,1123,242;%%


\bibitem[Piner (2008)]{}B.~G.~Piner, N.~Pant and P.~G.~Edwards,
%  %``The Parsec-Scale Jets of the TeV Blazars H 1426+428, 1ES 1959+650, and PKS
%  %2155-304: 2001-2004,''
  arXiv:0801.2749 [astro-ph].
%  %%CITATION = ARXIV:0801.2749;%%

\bibitem[Radomski, J.~T.(2008) et al.]{}Radomski, J.~T., et al.~2008,
Astrophys.\ J.\  {\bf 681}, 141R.

\bibitem[Reimer (2004)]{}Reimer, A, Protheroe, R.~J. and Donea, A.~C.,
  %``M87 as a misaligned synchrotron proton blazar,''
  New Astron.\ Rev.\  {\bf 48}, 411 (2004).
  %%CITATION = ASTRE,48,411;%%


\bibitem[Rejkuba, M.(2004)]{}Rejkuba, M.~2004, 
 Astron.\ Astrophys.\  {\bf 413}, 903.

\bibitem[Rieger (2008)]{}Rieger, F. M., \& Aharonian, F. A. 2008, Astron.\ Astrophys.\ {\bf 479}, L5.

\bibitem[Shakura, N.~I., \& Syunyaev, R.~A. (1973)]{}Shakura, N.~I., 
 Syunyaev, R.~A.~1973, Astron.\ Astrophys.\ {\bf 24}, 337.


\end{thebibliography}
\end{document}